\documentclass[review,11pt]{ReportTemplate}
\usepackage{bm}
\usepackage{graphicx}
\usepackage{paralist,algorithmic,algorithm}
\usepackage{amsmath,mathrsfs,amssymb}
\usepackage{hyperref}
\usepackage{subfigure}
\usepackage[export]{adjustbox}
\usepackage{geometry}
\usepackage{listings}
\usepackage{enumitem}
\usepackage[most]{tcolorbox}
\usepackage{xcolor}
\usepackage{booktabs}
\usepackage[capitalize,noabbrev]{cleveref}

\definecolor{codegreen}{rgb}{0,0.5,0}
\definecolor{codegray}{rgb}{0.5,0.5,0.5}
\definecolor{codepurple}{rgb}{0.58,0,0.82}
\definecolor{codeblue}{rgb}{0,0,0.5}
\definecolor{backcolour}{rgb}{0.95,0.95,0.92}

\lstdefinestyle{mystyle}{
    backgroundcolor=\color{gray!6},   
    commentstyle=\color{codegreen},
    keywordstyle=\bfseries\color{codeblue},
    numberstyle=\tiny\color{codegray},
    stringstyle=\color{codepurple},
    basicstyle=\ttfamily\footnotesize,
    breakatwhitespace=false,         
    breaklines=true,                 
    captionpos=b,                    
    keepspaces=true,                 
    numbers=left,                 
    showspaces=false,                
    showstringspaces=false,
    showtabs=false,                  
    tabsize=2,
    xleftmargin=0.01\textwidth,
    xrightmargin=0.01\textwidth
}
\newcommand{\bmwu}{\texttt{Beimingwu}}
\def \H {\mathcal{H}}
\def \s {\mathbf{s}}
\def \x {\mathbf{x}}
\def \z {\mathbf{z}}

\begin{document}
\begin{frontmatter}
\title{Beimingwu: A Learnware Dock System}
\renewcommand{\thefootnote}{\fnsymbol{footnote}}

\author{Zhi-Hao Tan\footnotemark[2], Jian-Dong Liu\footnotemark[2], Xiao-Dong Bi, Peng Tan, Qin-Cheng Zheng, \\
   Hai-Tian Liu, Yi Xie, Xiao-Chuan Zou, Yang Yu, Zhi-Hua Zhou\footnotemark[1]}
\address{National Key Laboratory for Novel Software Technology, Nanjing University, China\\
School of Artificial Intelligence, Nanjing University, China}

\begin{abstract} 
The learnware paradigm proposed by~\citet{Zhou2016} aims to enable users to reuse numerous existing well-trained models instead of building machine learning models from scratch, with the hope of solving new user tasks even beyond models' original purposes. In this paradigm, developers worldwide can submit their high-performing models spontaneously to the \textit{learnware dock system} (formerly known as \textit{learnware market}) without revealing their training data. Once the dock system accepts the model, it assigns a \textit{specification} and accommodates the model. This specification allows the model to be adequately identified and assembled to reuse according to future users' needs, even if they have no prior knowledge of the model. This paradigm greatly differs from the current \textit{big model} direction and it is expected that a learnware dock system housing millions or more high-performing models could offer excellent capabilities for both planned tasks where big models are applicable; and unplanned, specialized, data-sensitive scenarios where big models are not present or applicable.

\renewcommand{\thefootnote}{\fnsymbol{footnote}}
\footnotetext[1]{Corresponding author (email: zhouzh@lamda.nju.edu.cn)}
\footnotetext[2]{Equal contribution}

This paper describes \bmwu{}, the first open-source learnware dock system providing foundational support for future research of learnware paradigm.
The system significantly streamlines the model development for new user tasks, thanks to its integrated architecture and engine design, extensive engineering implementations and optimizations, and the integration of various algorithms for learnware identification and reuse. Notably, this is possible even for users with limited data and minimal expertise in machine learning, without compromising the raw data's security. \bmwu{} supports the entire process of learnware paradigm, including the submitting, usability testing, organization, identification, deployment and reuse of learnwares. 
The system lays the foundation for future research in learnware-related algorithms and systems, and prepares the ground for hosting a vast array of learnwares and establishing a learnware ecosystem.
\end{abstract} 
\end{frontmatter}

\section{Introduction}\label{sec:intro}

Numerous applications that rely on machine learning models have been integrated into many facets of modern life. However, in classic machine learning paradigm, to train a high-performing model from scratch for a new task still requires an abundance of high-quality data, expert experience and computational resources, which is difficult and expensive.
There are also lots of concerns when reusing existing efforts, such as the difficulty of adapting a specific trained model to different environments, and the embarrassment of catastrophic forgetting when refining a trained model incrementally. Besides, privacy and proprietary issues hinder the data sharing among developers, and restrict the capabilities of big models in many data-sensitive scenarios. 
Indeed, most efforts have been focusing on one of these concerned issues separately, paying less attention to the fact that most issues are entangled in practice.

On the other hand, the prevailing big model paradigm, which has achieved successful results in natural language processing~\citep{Brown:Mann:Ryder:Subbiah2020} and computer vision~\citep{Radford:Kim:Hallacy:Ramesh2021} applications, has not yet addressed the above issues. It would be too ambitious to build a big model for every possible task, because of the infinite nature of unplanned tasks and scenarios, constantly changing environments, catastrophic forgetting, excessively high demand for resources, privacy concerns and local deployment requirements, and personalized and customized demands.

To tackle the above issues simultaneously and leverage existing efforts in a systematic and unified way, \textit{learnware}~\citep{Zhou2016,Zhou:Tan2024} was proposed, based on which machine learning tasks can be solved in a novel paradigm. 
For the first time, the learnware paradigm proposes to build a foundational platform, i.e., \emph{learnware dock system}, which uniformly accommodates numerous machine learning models submitted spontaneously by developers worldwide, then leverages the capabilities of numerous models to solve new tasks according to future users' needs. 
The core design of the paradigm is as follows: 
For high-performing models of any structure from various tasks, a learnware is a uniformly formatted fundamental unit, consisting of the model itself and a \emph{specification} which captures the model's specialty in a certain representation, and enables it to be adequately identified and reused according to the requirement of future users.
A learnware dock system, serving as the foundation of the paradigm, generates unified specifications for submitted models. By accommodating all learnwares, when facing a emerging new user task, the learnware dock system will identify and assemble helpful learnware(s) for the user based on specifications. Then these learnware(s) can be applied by the user directly or adapted by her own data for better usage. 
Besides, the learnware dock system should be able to protect the the original training data of model developers and users.

To establish the foundation for learnware paradigm, we built \bmwu{}, the first open-source learnware dock system for future research of learnware paradigm. 
Benefiting from scalable system and engine architecture design, extensive engineering implementations and optimizations, integration of baseline algorithms for the entire process, and construction of convenient algorithm evaluation scenarios, the system not only provides a foundation for future research in learnware-related algorithms and dock system studies, but also lays the groundwork for hosting a vast array of learnwares and establishing a learnware ecosystem. In this paper, our contributions can be summarized as follows:

\begin{itemize}[leftmargin=*]
  \item {\bf Streamlining the model development for new tasks: data-efficient, expert-free and privacy-preserving.}  
  Based on the first systematic implementation of learnware paradigm, \bmwu{} significantly streamlines the process of building machine learning models for new tasks, and we can now construct models following the process of learnware paradigm. 
  The following exciting advancements become possible: If there exist some learnwares in \bmwu{} possessing capabilities helpful for the task, one can utilize \bmwu{} to obtain and deploy a high-performing model with just a few lines of code, without extensive data and expert knowledge, while ensuring data privacy.

  \item {\bf Integrated and scalable architecture design for system engine.}
  To solve new tasks in learnware paradigm, a learnware dock system should be able to accommodate and leverage submitted high-performing models from various tasks in a unified way. 
  In \bmwu, we specify a unified learnware structure, and design an integrated architecture for system engine, which can support the entire process including the submitting, usability testing, organization, management, identification, deployment and reuse of learnwares. 
  The architecture is scalable to coordinate a vast array of learnwares, and possesses unified and scalable interfaces for future research in organization, identification and reuse algorithms.

  \item {\bf An open-source learnware dock system with unified user interface.}
  Based on the engine architecture and reduced kernel mean embedding (RKME) specification~\citep{Zhou:Tan2024}, through extensive implementations and refining, we develop the core engine of \bmwu{} and release as \texttt{learnware} package, which supports the computational and algorithmic aspects of \bmwu{}.
  Furthermore, through extensive designing of system architecture and engineering optimizations, we develop the system backend and user interface including web frontend and command-line client for better establishment of learnware ecosystem. \bmwu{} is open-source for collaborative community contributions and convenience of learnware research.

  \item {\bf Implementation and evaluation of full-process baseline algorithms for various scenarios.} 
  For specification generation, learnware organization, identification and reuse, we have implemented and refined a set of baseline algorithms to support models trained on tabular, image, and text data. These algorithms enable the identification and reuse of both single and multiple learnwares.
  We also provide a series of baseline algorithms to support the organization, identification and reuse of learnwares from different feature spaces.
  Besides, we build various types of experimental scenarios and conduct corresponding empirical study for evaluation, which are all public for future research.

\end{itemize}

{\bf Organization.} We start with a brief review of learnware paradigm in Section~\ref{sec:review}, then we introduce how to solve a new learning task by using \bmwu{} in Section~\ref{sec:solving}. Section~\ref{sec:system} presents the overview of the entire system architecture, and the architecture design of the core engine, following by the introduction of baseline algorithms. We conduct empirical study to evaluate the baseline algorithms on various types of experimental scenarios in Section~\ref{sec:evaluation}. Finally we discuss the related works and conclude the paper.

\section{A Brief Review of Learnware Paradigm}
\label{sec:review}
The learnware paradigm was proposed in~\citet{Zhou2016}. A learnware is a well-performing machine learning model with a \textit{specification} which captures the model's specialty in a certain representation, and enables it to be adequately identified to reuse according to the requirement of future users.

\begin{figure}[t]
\centering
\includegraphics[width=\textwidth]{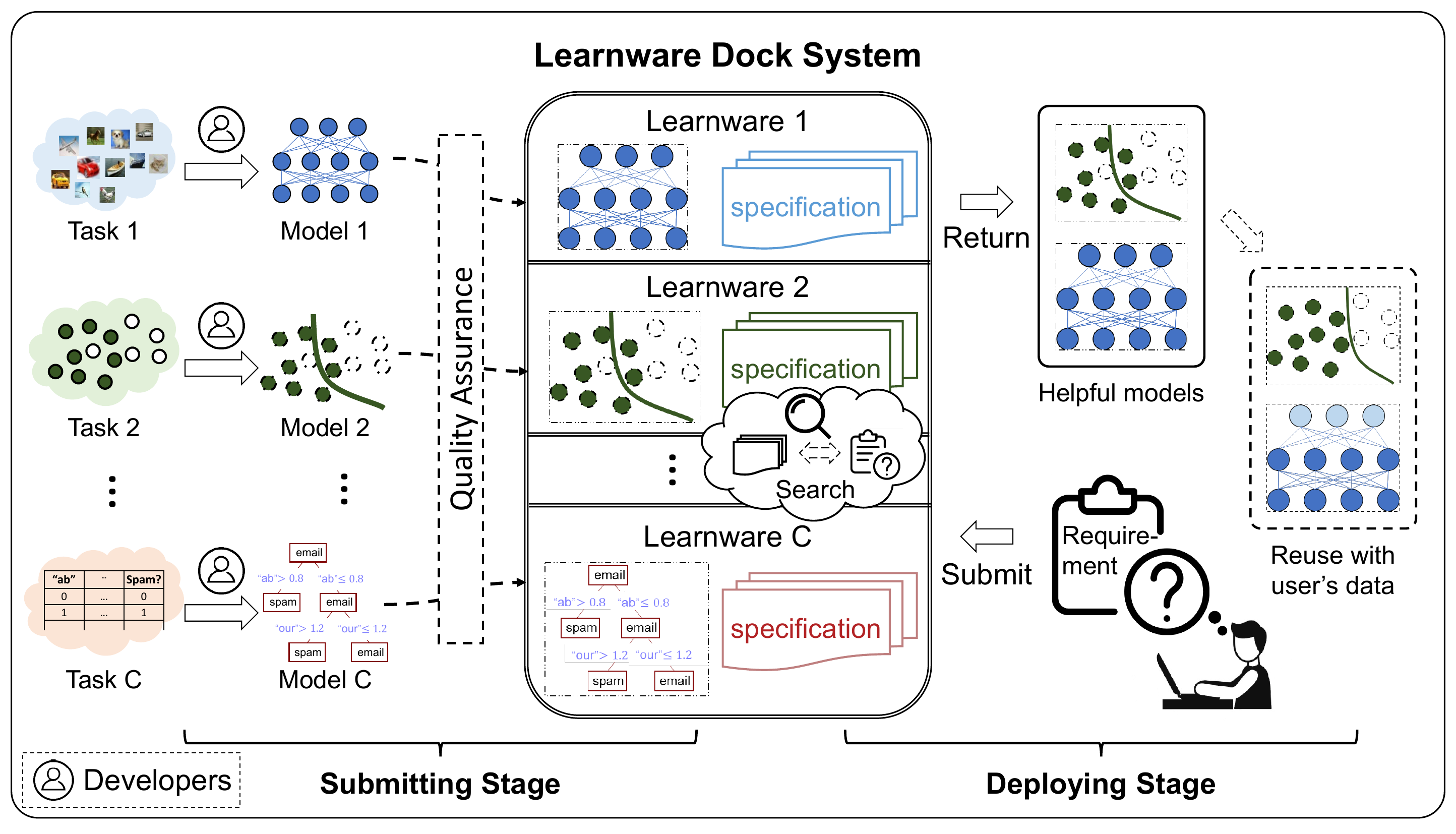}
\caption{A simplified process of learnware paradigm from~\citet{Zhou:Tan2024}. The basic operation can be decomposed into two stages:
1) {\bf Submitting stage}: Developers worldwide can spontaneously submit their trained models to the learnware dock system, and the system assigns specification for each accepted model;
2) {\bf Deploying stage}: The user submits her requirement to the learnware dock system, and then the system will identify and return some helpful learnwares to the user based on specifications, which can be further reused on user data.
}
\label{fig:stages}
\end{figure}

Recently, the learnware paradigm was summarized and further designed in~\citet{Zhou:Tan2024}, where the simplified process of the paradigm is illustrated in~\cref{fig:stages} and outlined as follows:
The developer or owner of a well-performing machine learning model, of any type and structure, can spontaneously submit her trained model into a \textit{learnware dock system} (previously called \textit{learnware market}). If the model passes the quality detection, the learnware dock system will assign a specification to the model and accommodate it in the system in a unified way. 
The learnware dock system should be scalable to accommodate thousands or millions of well-performed models submitted by different developers, on different tasks, using different data, optimizing different objectives, etc.

Based on the learnware dock system, when a user is going to solve a new machine learning task, she can submit her \textit{requirement} to the learnware dock system, and then the system will identify and assemble some helpful learnware(s) from numerous learnwares to return to the user by considering the learnware specification. She can apply the learnware(s) directly, or adapt them by her own data, or exploit in other ways to help improve the model built from her own data.
No matter which learnware reuse mechanism is adopted, the whole process can be much less expensive and more efficient than building a model from scratch by herself. Importantly, it has been proved that the learnware paradigm has privacy-preserving ability, which enables developers to share their models that can be adequately identified and reused by future users without disclosing developer's original training data.

The learnware paradigm proposes to build a learnware dock system to accommodate, organize and leverage existing well-performing models uniformly, which provides a unified way to leverage existing efforts from all the community to solve new user tasks, and offers the possibility of addressing significant concerned issues simultaneously~\citep{Zhou:Tan2024}: Lack of training data, Lack of training skills, Catastrophic forgetting, Hard to achieve continual learning, Data privacy/proprietary, Unplanned new tasks from open world, and Carbon emissions caused by wasteful repetitive training. 

Recently the learnware paradigm and its key idea have received increasing attention.
The key problem and main challenge is that, considering a learnware dock system which has accommodated thousands even millions of models, how to identify and select the most helpful learnware(s) for a new user task? Apparently, direct submitting user data to the system for trials would be unaffordable and leak user's raw data.
The core design of learnware paradigm lies in the specification.
Recent progresses are mainly based on the reduced kernel mean embedding (RKME) specification~\citep{Zhou:Tan2024}.
For example,~\citet{Wu:Xu:Liu:Zhou2023} proposed to identify helpful learnware(s) by matching original data distributions of learnwares with user data distributions based on RKMEs, which is further extended by~\citet{Zhang:Yan:Zhao:Zhou2021} under the existence of unseen parts in user tasks. By learning a unified specification space for various learnwares from heterogeneous feature spaces,~\citet{Tan:Tan:Jiang:Zhou2022,Tan:Tan:Jiang:Zhou2023} proposed learnware search and reuse algorithms to utilize the learnware from heterogeneous feature spaces. To support efficient and accurate identification from a large number of learnwares,~\citet{Xie:Tan:Jiang:Zhou2023} 
proposed the anchor-based mechanism, which organizes learnwares structurally and identifies helpful learnwares by only accessing a small number of anchor learnwares instead of examining all learnwares.~\citet{Liu:Tan:Zhou2024} further proposed an efficient learnware identification method called Evolvable Learnware Specification with Index (ELSI), resulting in increasingly accurate characterization of model abilities beyond original training tasks with the ever-increasing number of learnwares. Besides,~\citet{Guo:Zhou:Li:Zhou2023} attempted to leverage learnwares from heterogeneous label spaces. 

Although existing study in theoretical and empirical analysis has shown the effectiveness of specification-based learnware identification of the learnware paradigm, the realization of a learnware dock system is still missing and remains a big challenge, which needs a novel specification-based architecture design to handle the diversity of real-world tasks and models, and to leverage numerous learnwares in a unified way according to the user task requirement. In this paper, we built \bmwu, the first learnware dock system, which can support the entire process including the submitting, usability testing, organization, management, identification, deployment and reuse of learnwares. Based on \bmwu{}, we lay the groundwork for future research and ecosystem, and show the effectiveness of learnware paradigm in addressing concerned issues in Section~\ref{sec:intro} simultaneously.

\section{Solving Learning Tasks with \bmwu}
\label{sec:solving}

Based on the first systematic implementation of learnware paradigm, \bmwu{} significantly streamlines the process of building machine learning models for new tasks, and we can now construct models following the process of learnware paradigm. 
Benefiting from specifying a unified learnware structure, integrated architecture design and unified user interfaces, all submitted learnwares in \bmwu{} can be uniformly identified and reused.

Excitingly, given a new user task, if \bmwu{} possesses learnwares with capabilities to tackle the task, with just a few lines of code, the user can easily obtain and deploy a high-quality model based on \bmwu{}, without requiring extensive data and expert knowledge, while not revealing her original data, as presented in~\cref{fig:example}.

\begin{figure}[t]
\centering
\begin{lstlisting}[language=Python]
from learnware.market import BaseUserInfo
from learnware.specification import generate_stat_spec
from learnware.client import LearnwareClient
from learnware.reuse import AveragingReuser

# Generate statistical specification for local data
rkme = generate_stat_spec(type="table", X=data)
user_info = BaseUserInfo(stat_info={rkme.type: rkme})

# Beimingwu identifies single or multiple helpful learnwares
learnware_ids = client.search_learnware(user_info)["multiple"]["learnware_ids"]

# Load the returned learnwares uniformly
learnware_list = client.load_learnware(learnware_id=learnware_ids, runnable_option="docker")

# Reuse learnwares on own data
y_pred = AveragingReuser(learnware_list, mode="vote_by_label").predict(data)
\end{lstlisting}
\caption{Practical codes for solving a learning task with \bmwu{}. With just a few lines of code, a user can build a model for her limited data with the help of numerous learnwares in \bmwu{}, without requiring extensive data and machine learning expertise, while not leaking her raw data.
}\label{fig:example}
\end{figure}

Users can also access the system with web frontend to easily interact with the system to see information of each learnware and to obtain helpful learnwares by choosing semantic specification, like data type, task type and scenarios, or uploading statistical RKME specification for precise identification. Note that the core engine of \bmwu{}, supporting the computational and algorithmic aspects, are extracted and released as \texttt{learnware} package, which can be easily used locally. 

\begin{figure}[!t]
    \centering
    \includegraphics[width=\linewidth]{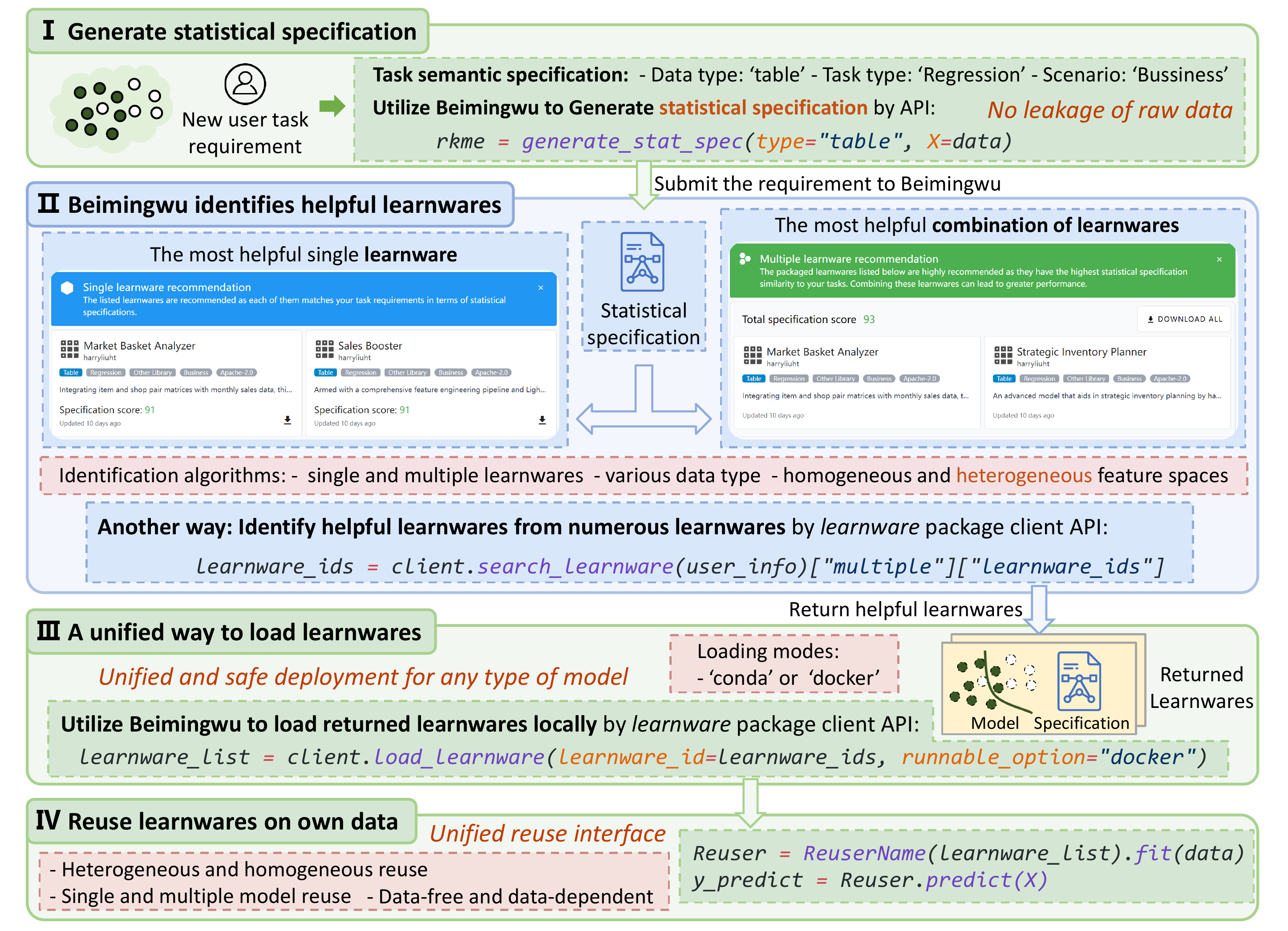}
    \caption{
        Overview of using \bmwu{} to solve new learning tasks. The workflow consists of four steps: 
        1) {\bf Generating statistical specification}: \bmwu{} helps the user to generate statistical specification capturing the statistical property of the task without disclosing user's raw data; 
        2) {\bf Identify helpful learnwares}: According to the submitted task requirement, \bmwu{} can identify helpful learnware(s) from numerous learnwares for the user based on learnware specifications;
        3) {\bf Loading learnwares}: \bmwu{} provides a unified way to load arbitrary learnwares effortlessly and safely;
        4) {\bf Reuse learnwares}: \bmwu{} provides various baseline reuse algorithms in a unified interface to reuse learnwares on user data. 
    }\label{fig:using_bmwu}
\end{figure}

The entire workflow of using \bmwu{} is shown in Figure~\ref{fig:using_bmwu}, including statistical specification generation, and learnware identification, loading and reuse. Based on extensive engineering implementations and unified interface design, each step can be achieved with one key line of code.
Note that the statistical specification is realized via reduced kernel mean embedding (RKME) specification~\citep{Zhou:Tan2024} which captures the data distribution while not disclosing the raw data. Detailed learnware identification algorithms based on statistical specifications and learnware reuse algorithms will be presented in Section~\ref{subsec:algorithm}.

As envisioned in Section~\ref{sec:review}, based on \bmwu{} system, we demonstrate that the learnware paradigm offers a promising way to solve the issues mentioned in Section~\ref{sec:intro}. Specifically, the model development process for a learning task based on \bmwu{} possesses the following significant advantages:
\begin{itemize}[leftmargin=*]
  \item {\bf No need for extensive data and training resources.} If helpful learnwares exist, \bmwu{} identifies and assembles helpful learnware(s) for user tasks from numerous learnwares in the system, then users can directly utilize them or refine them with a small amount of data, instead of training a new model with extensive data and resources from scratch.

  \item {\bf Minimal machine learning expertise.}  With just a few lines of code, users can easily obtain suitable learnware(s) identified by the system for their specific tasks. This streamlined process makes numerous high-quality and potentially helpful models adequately utilized by users across all levels of expertise. It eliminates the need for expert knowledge in designing priors, or manually selecting algorithms and models. 

  \item {\bf Simple and secure local deployment of diverse models.} 
  Ideally, the system accommodates a variety of high-quality learnwares submitted by developers worldwide, applicable to a wide range of specialized and customized scenarios. By identifying and reusing the most suitable small and deployable learnwares for specific tasks, local deployment can mostly be achieved, offering a practical alternative to relying on a single cloud-based big model, especially for unplanned, specialized, data-sensitive scenarios.
  Besides, based on engineering implementations and architecture optimizations, and specifying a unified learnware structure, \bmwu{} allows for effortless and safe deployment and reuse of arbitrary learnwares in a unified way based on containerized isolation, with little concerns about environment compatibility and safety.

  \item {\bf Privacy-preserving: no leakage of original data.} To identify the most suitable learnwares, instead of uploading the original data, a user generates and submits the RKME statistical specification to the system using API, which captures the data distribution while not disclosing the raw data. Based on the RKME, \bmwu{} identifies the learnwares that are most beneficial for user task and returns them to the user. More importantly, the privacy-preserving ability enables developers to share their models that can be adequately identified and reused by future users without disclosing developer’s original training data.
\end{itemize}

Presently, at the initial stage, \bmwu{} houses only about 1100 learnwares built from open-source datasets covering limited scenarios, which still has limited capabilities for numerous specific and unforeseen scenarios. 
However, by implementing the integrated specification-based architecture, we have specified a unified user interface such that user can easily obtain helpful learnwares identified by the system and deploy any learnware uniformly; and based on the scalable architecture, \bmwu{} also serves as a research platform for learnware paradigm, which supports convenient algorithm implementation and experimental design for learnware-related studies.

Besides, relying on the foundational implementations and scalable architecture, the constantly submitted learnwares and algorithmic advancement will expand the knowledge base of the system and enhance its ability to reuse existing well-trained models to solve new user tasks even beyond their original purposes, and this continuous evolution of learnware dock system enables it to respond to more and more user tasks without catastrophic forgetting, naturally realizing lifelong learning.

\section{The Design of \bmwu{}}
\label{sec:system}

\begin{figure*}[t]
    \centering
    \includegraphics[width=\linewidth]{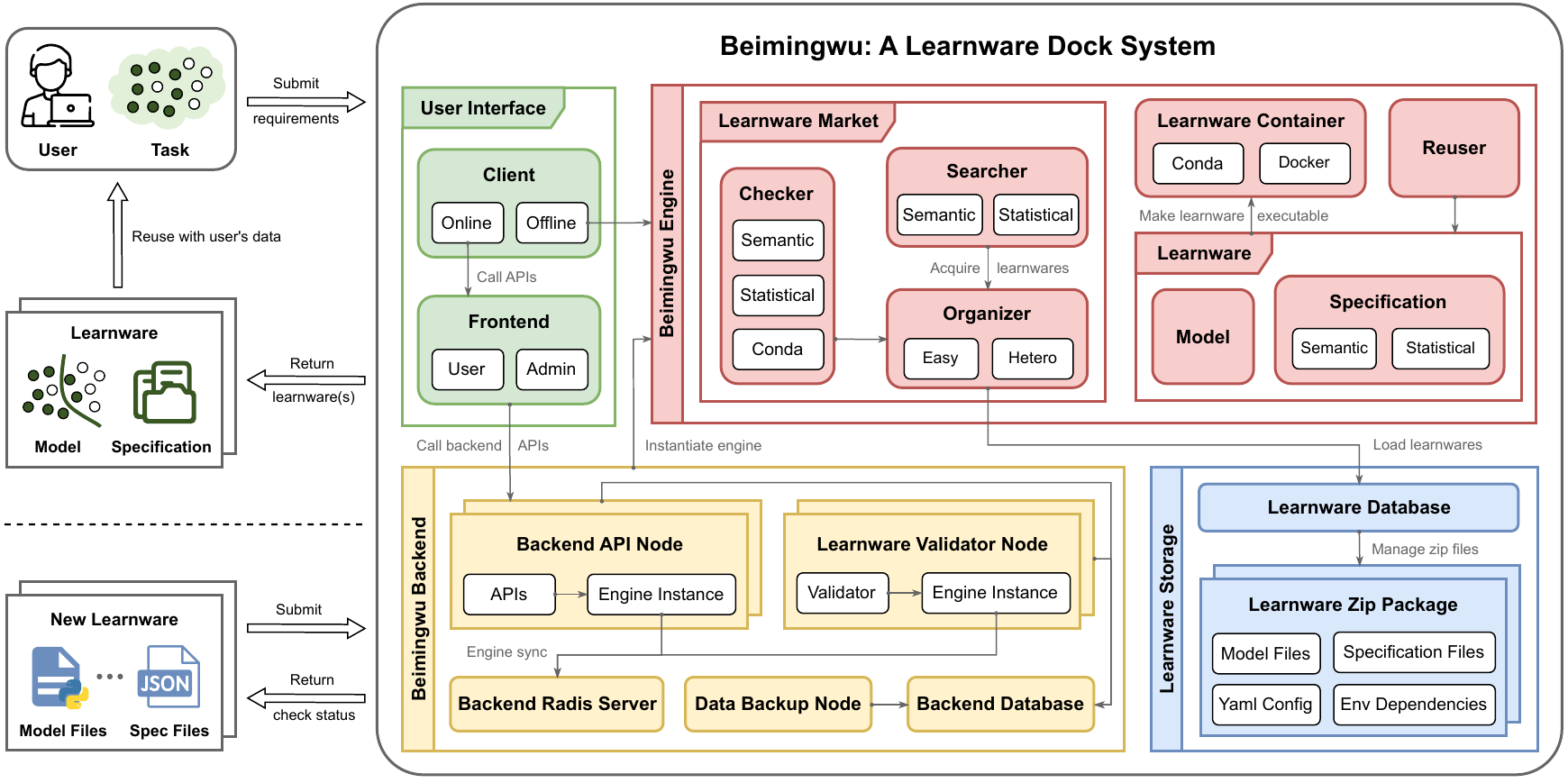}
    \caption{
        Architecture of \bmwu{}.
    }\label{fig:system-arch}
\end{figure*}

In this section, we will introduce the design of \bmwu{} system. 
As briefly depicted in~\cref{fig:system-arch}, the entire system comprises four hierarchical layers: learnware storage, system engine, system backend, and user interface. In the following subsections, we will first provide an overview of each layer, then present our specification-based design of core engine of the system, finally introduce the algorithms implemented in the system.

\subsection{Overview of \bmwu{} Architecture}

{\bf Learnware storage layer.}
In \bmwu{}, learnwares are stored in the form of zip packages, with their specific format defined based on the architecture outlined in Section~\ref{subsec:arch-engine}. These packages primarily comprise four types of files: model files, specification files, environment dependency files for model execution, and learnware configuration files.

These learnware zip packages are centrally managed by the learnware database. The learnware table within this database stores crucial information, including the learnware ID, storage path, and learnware status (e.g., unverified and verified). This database provides a unified interface for the subsequent core engine of \bmwu{} to access learnware information.

Additionally, the database can be constructed using either SQLite (suitable for easy setup in development and experimental environments) or PostgreSQL (recommended for stable deployment in production environments), both utilizing the same management interface.

{\bf Core engine layer.}
To maintain the simplicity and structure of \bmwu{}, we have separated the core components and algorithms from the extensive engineering details. These extracted elements are now available as \texttt{learnware} package, which serves as the core engine of \bmwu{}.

As the system kernel, the engine encompasses all processes within the learnware paradigm, including the submitting, usability testing, organization, identification, deployment and reuse of learnwares. It operates independently of the backend and frontend, offering rich algorithmic interfaces for learnware-related tasks and research experiments.

Moreover, characterizing the associated model from both semantic and statistical perspectives, \textit{specification} serves as the central component in the engine, connecting various essential learnware-related components.
In addition to users' regular specifications, the engine leverages the system's knowledge to generate new system specifications for learnwares, enhancing learnware management and further characterizing their capabilities.

In contrast to existing model management platforms, such as the Hugging Face Hub, which passively collect and store models, leaving users to determine model capabilities and relevance to their tasks, \bmwu{} actively manages learnwares through its engine. This active management goes beyond collection and storage. The system organizes learnwares based on specifications, proactively recommends relevant learnwares to users, and provides corresponding methods for learnware reuse and deployment.

{\bf System backend layer.}
To enable industrial-level deployment of \bmwu{}, we have developed the system backend, building upon the core engine layer. Through the design of multiple modules and extensive engineering development, \bmwu{} is now capable of online and stable deployment, providing comprehensive backend APIs to the frontend and clients.

To ensure efficient and stable system operation, we have implemented several engineering optimizations in the system backend layer:
\begin{itemize}[leftmargin=*]
    \item \textbf{Asynchronous learnware validation.} Synchronous validation responses to concurrent learnware uploads can severely impact system efficiency. To address this challenge and efficiently validate uploaded learnwares, we have designed and developed a learnware validator node for asynchronous validation. This node continuously retrieves learnwares with a "waiting" status from the database and processes them in a queue, significantly reducing system stress.
    \item \textbf{High concurrency across multiple backend nodes}: Both the API node and validator node in the backend run concurrently, with each node's engine instance synchronized through a Redis server. When an engine instance completes a learnware-related write operation, it sends a message to other engine instances via the Redis server. The other engine instances then reload the corresponding learnware, ensuring synchronization across nodes.
    \item \textbf{Interface-level permission management}: To facilitate effective system management and convenient usage, we have designed different permissions for various backend APIs, including: no login required, login required, and administrator permissions required.
    \item \textbf{Backend database read-write separation}: To improve the overall system efficiency, \bmwu{} adopts a master-slave database architecture, implementing read-write separation for the backend database. This means that database read and write operations are distributed to different database instances, enhancing the efficiency of various database operations.
    \item \textbf{Automatic system data backup}: To enhance system security, we have designed a data backup node that automatically backs up system data at regular intervals, ensuring the safety of user and learnware information.
\end{itemize}

{\bf User interface layer.}
For the convenience of \bmwu{} users, we have developed the corresponding user interface layer, including a web-based frontend for browser access and a command-line client.

The web-based frontend serves both user and administrative requirements, providing a diverse range of user interaction and system management pages. Additionally, it supports multi-node deployment for smooth access to \bmwu{}.

The command-line client is seamlessly integrated into the \texttt{learnware} package. Through this client, users can invoke backend online APIs via the frontend and access learnware-related modules and algorithms packaged in \texttt{learnware}.

\subsection{Engine Architecture Design}
\label{subsec:arch-engine}

\begin{figure}[!t]
\centering
\includegraphics[width=0.95\textwidth]{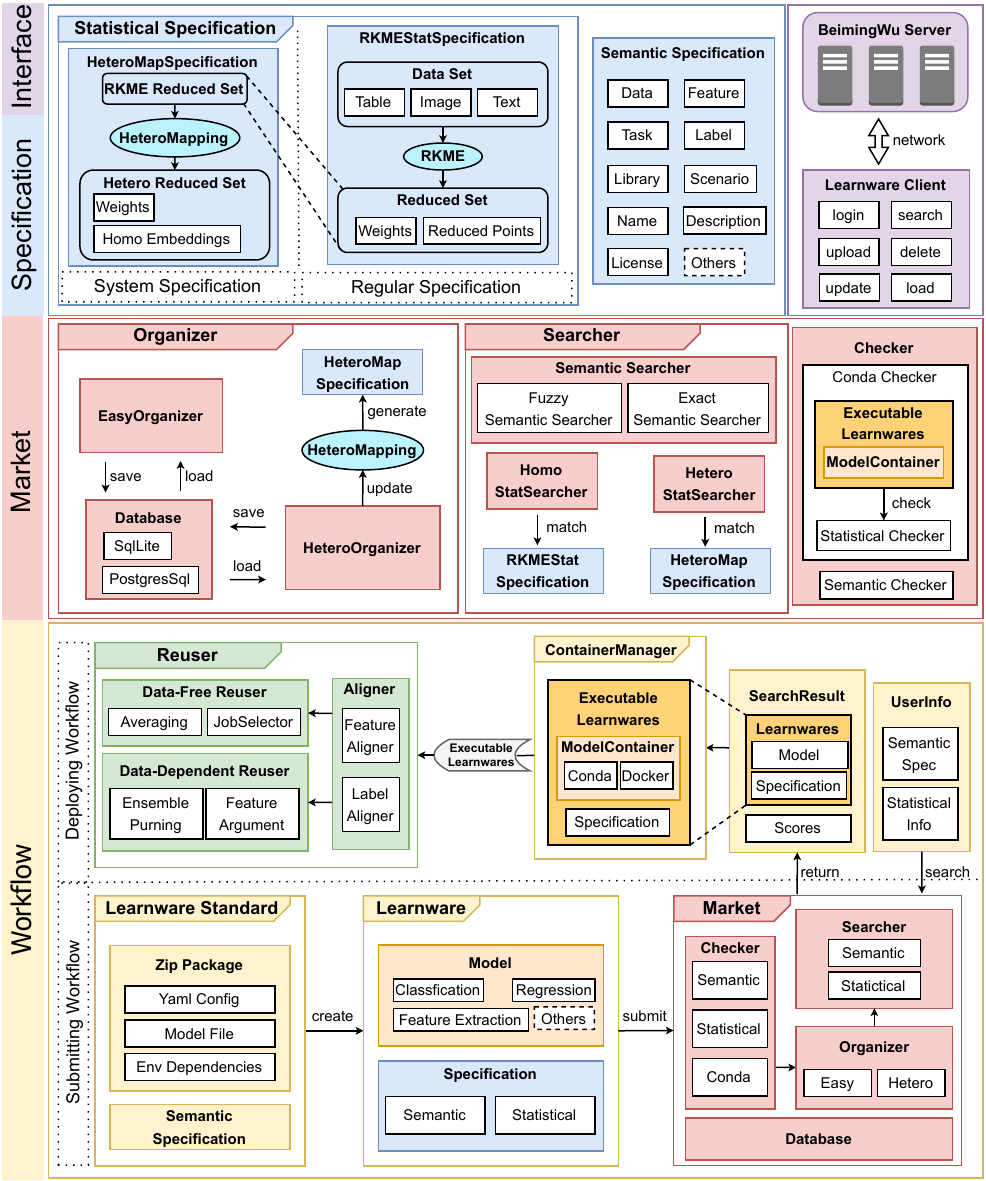}
\caption{Architecture design of \bmwu{} engine. The architecture is illustrated from the perspectives of both modules and workflow.}
\label{fig:engine}
\end{figure}

In this section, we present the infrastructure design of the \texttt{Beimingwu} engine. We begin by outlining the design principles, followed by a description of the engine architecture from both the perspectives of modules and processes.

\subsubsection{Design Principles}

We design the \texttt{Beimingwu} engine based on the following guidelines that include decoupling, autonomy, reusability, and scalability.

\noindent\textbf{Decoupling}. The engine should be decoupled from the backend of the \texttt{Beimingwu} system. The engine itself should support academic algorithms and learnware processes, including submission, usability testing, organization, management, identification, deployment, and reuse of learnwares. On the other hand, the backend is responsible for business logic such as user login, registration, user information management, as well as engineering implementations like flow monitoring, multi-process concurrency, and containerized isolation. This separation ensures that the modules related to academic algorithms and learnware processes are decoupled from business-related and engineering modules, making the \texttt{Beimingwu} system less cluttered and easier to maintain.

\noindent\textbf{Autonomy}. In addition to serving as an integral part of the system to undertake core functions, the engine should also be capable of functioning as an independent academic experimental platform. As an autonomous entity, the engine should not only encompass all submission and deployment processes of learnware but also provide interfaces for interaction. This way, in future academic algorithm research, users can test their novel algorithms by directly modifying and interacting with the engine rather than the entire \texttt{Beimingwu} system, which contains too much engineering implementation unrelated to algorithms and is more challenging to deploy compared to the engine.

\noindent\textbf{Reusability}. The engine will integrate a large number of learnware algorithms at different stages, including algorithms for learnware organization, identification, usability testing, and the reuse of learnwares. Therefore, the engine architecture needs to be designed in a way that each module is responsible for an appropriate amount of functionality. This architecture should promote the reusability of code in each module, enabling the combination of different modules to create diverse learnware processes. For example, users should be able to easily achieve various learnware identification and reuse processes by combining different modules.

\noindent\textbf{Scalability}. With the progress of learnware algorithm research, more algorithms will be integrated into the engine. Therefore, the engine architecture should be highly scalable. Each module within the engine should provide interfaces for extending new methods, thus supporting more powerful learnware organization, identification, and reuse algorithms in the future. This also allows users to easily integrate their novel algorithms into the academic experimental platform to test the performance of these new algorithms.

\subsubsection{Design of Core Modules}

According to the design principles, we have created various modules for the engine, encompassing the learnware module, market module, specification module, model module, reuse module, and interface module. The corresponding modules are depicted in~\cref{fig:engine}.

\noindent\textbf{Learnware}. The learnware module represents specific learnware within the dock system, comprising a learnware identifier, a specification module, and a user model module. The dock system can generate a learnware module by parsing learnware files stored in accordance with the learnware standard format. As depicted in Figure 1, the learnware standard includes learnware configuration files, model files, statistical specification files, semantic specification files, and environment configuration files. Specifically, (a) the learnware configuration file assists the dock system in parsing and instantiating the learnware module. (b) The model file details the user model, providing a unified interface for training, prediction, and fine-tuning. (c) The statistical specification file is utilized to generate and store statistics related to statistical specifications. (d) The semantic specification file outlines task types, data types, scenarios, and other semantic information of the learnware. (e) The environment configuration file describes the software dependencies of user models.

\noindent\textbf{Market}. The market module is designed for learnware organization, identification, and usability testing. A single market module consists of one organizer, one searcher module, and multiple checkers. (a) The organizer module oversees the storage and organization of learnware, supporting operations such as reloading the entire learnware collection and performing insertions, deletions, and updates. These operations are currently implemented using database techniques. (b) The searcher module conducts learnware identification based on statistical and semantic specifications. It filters learnware based on semantic specifications and matches learnware by comparing the similarity between user-provided statistical specifications and the statistical specifications of the learnware itself. (c) The checker module is responsible for checking the usability and quality of learnwares by verifying the availability of semantic and statistical specifications and creating a runtime environment to test user models based on the model container.

\noindent\textbf{Specification}. The specification module is employed to generate and store statistical and semantic information for learnware, which is utilized in learnware search. The specification module comprises both semantic and statistical specification modules. Within the statistical specification module, there are both system statistical specifications and user statistical specification modules. (a) The user statistical specification module is locally generated by the user and uploaded to the system. Currently, it includes statistical specification modules based on RKME (Reduced Kernel Mean Embedding), applicable to tabular, image, and text data. The same type of RKME statistical specification modules can be evaluated for similarity using Maximum Mean Discrepancy (MMD) calculations. During learnware search, the dock system matches and recommends learnware based on the similarity of statistical specification modules. (b) The system statistical specification module is automatically generated by the dock system. For newly inserted learnware, the organizer generates new system specifications based on existing user statistical specifications to facilitate search operations. Currently, the dock system has implemented the heterogeneous mapping specification for the system statistical specification module. When inserting learnware with heterogeneous tabular data, the learnware organization module maps the user-generated RKME specifications to the same embedding space with heterogeneous mapping to support search operations on heterogeneous tabular data. For specific details about heterogeneous search, please refer to Section~\ref{subsec:algorithm}. 

\noindent\textbf{Model}. The model module comprises the base model interface and the model container. (a) The base model interface is crafted to standardize interfaces for model training, prediction, and fine-tuning. All user models inherit from this interface, allowing the system to invoke user models for all learnware based on a unified interface. (b) The model container, also inherited from the base model interface, automatically establishes an isolated runtime environment based on the environment configuration file and executes user models within that environment.

\noindent\textbf{Reuse}. The reuse module comprises the data-free reuser, data-dependent reuser, and aligner. Currently, the data-free reuser derives the final prediction directly, employing methods such as ensemble models from multiple learnwares. The data-dependent reuser, utilizing additional labeled data provided by users, obtains final prediction results through methods such as fine-tuning a meta-model. Both methods require that the features and prediction spaces of the input learnwares are in the same dimension. The aligner maps the features and prediction spaces of heterogeneous learnwares to the same space. To handle learnwares with different dimensions, the dock system passes them through the aligner to obtain feature-aligned learnwares, which can then be passed to either the data-free reuser or data-dependent reuser for reuse. For specific details regarding the implementation of the aligner, please refer to Section~\ref{subsec:algorithm}.

\noindent\textbf{Interface}. The engine also integrates an interface for network communication with the backend. Upon installing the engine, users can leverage this interface via the command line to effortlessly and efficiently execute batch operations. These operations encompass swift uploading, deletion, and updating of learnwares within the deployed learnware dock system on the system server.

\subsubsection{Submitting and Deploying Processes}

To provide a clearer insight into the data flow and operational procedures of the engine architecture, we delve into the submission and deployment stages of learnwares. This section refers to the processes and various modules illustrated in~\cref{fig:engine} for a more comprehensive understanding.

\noindent\textbf{Submitting process}. 
During the submission stage, learnware developers submit learnwares to the learnware dock system, where the dock system conducts usability checks and organizes these learnwares. Now, we illustrate the detailed process for submitting learnwares in the engine. (a) The learnware developers generate learnware storage zip packages in accordance with the standard learnware format and upload them to the system. (b) The dock system parses the uploaded learning storage zip packages based on the learnware configuration file, generating learnware modules. (c) The instantiated learnware modules are uploaded to the checker module for usability checks. (d) The learnwares that pass the check are then inserted and stored by the organizer module.

\noindent\textbf{Deploying process}. During the deployment stage, the dock system recommends learnwares based on users' task requirements, offering efficient reuse and deployment methods. Now, we present the detailed process for deploying learnwares in the engine. (a) The learnware users generate user information, comprising statistical and semantic specification modules, based on their machine learning tasks and requirements. (b) The learnware users upload the user information to the searcher within the market module. The searcher module employs a search algorithm to retrieve learnwares that meet user requirements and recommends them as search results. (c) The docker system uses the model container to containerize the learnware modules in the search results and deploys the learnwares in the runtime environments. (d) The containerized learnware modules are passed to the aligner, which returns learnware modules with aligned features and prediction spaces. (e) The feature-aligned learnware modules are then passed to either the data-free reuser or the data-dependent reuser for reuse.

\subsection{Implemented Algorithms of Beimingwu System}
\label{subsec:algorithm}

{\bf System specifications.} Specifications implemented by the system are all derived from the RKME specification~\citep{Zhou:Tan2024}, which uses techniques based on the reduced set of KME (Kernel Mean Embedding)~\citep{Scholkopf:Mika:Burges:Knirsch:Muller1999,Berlinet:Thomas-Agnan2011}. Suppose a developer is to submit a model trained from data set $\{(\x_i, y_i)\}_{i=1}^m$, the reduced set representation $\{(\beta_j,\z_j)\}_{j=1}^n$ is generated by minimizing the distance between the KMEs of the reduced set and the original data measured by the RKHS norm:
\begin{align}
    \min _{\boldsymbol{\beta}, \mathbf{Z}}\left\|\frac{1}{m} \sum_{i=1}^m k\left(\x_i, \cdot\right)-\sum_{j=1}^n \beta_j k\left(\z_j, \cdot\right)\right\|_{\mathcal{H}}^2,
\end{align}
with non-negative constraints of coefficients $\{\beta_j\}_{j=1}^n$. The RKME $\Phi(\cdot) = \sum_{j=1}^n \beta_j k(\z_j, \cdot) \in \H$ serves as the specification which offers a concise representation of the original data distribution $\mathcal{P}$ while preserving data privacy. 

The Beimingwu system provides a unified interface \texttt{generate\_stat\_spec}  for model developers and users to easily generate specifications for specific data types. This generation process runs locally and does not involve any data sharing with the backend system, thus ensuring data privacy and ownership. The system supports specifications for various data types:

\begin{itemize}[leftmargin=*]
    \item \textbf{Tabular specification:} For tabular tasks, the system generates two distinct types of specifications. The first,  \texttt{RKMETableSpecification}, implements the RKME specification, which is the basis of tabular learnwares. It facilitates learnware recommendation and reuse for homogeneous tasks with identical input and output domains. The second, \texttt{HeteroMapTableSpecification},   enables learnware to support tasks with varying input domains and output domains. This specification is derived from \texttt{RKMETableSpecification} and is produced using the system's heterogeneous engine. This engine, a tabular network, is trained on feature semantics of all tabular learnwares in the system. 

    \item \textbf{Image specification:} The specification for image data \texttt{RKMEImageSpecification} introduces a new kernel function that transforms images implicitly before RKME calculation. It employs the Neural Tangent Kernel (NTK)~\citep{Garriga-Alonso:Aitchison:Rasmussen2019}, a theoretical tool that characterizes the training dynamics of deep neural networks in the infinite width limit, to enhance the measurement of image similarity in high-dimensional spaces.

    \item \textbf{Text specification:} Unlike tabular data, text inputs of varying lengths are processed into sentence embeddings using multilingual embedding models. Subsequently, the RKME specification \texttt{RKMETextSpecification} is calculated based on these embeddings. 
\end{itemize}

{\bf System search algorithms.} When the user submits her task requirement, the system searches useful learnwares through leveraging learnware specifications and the user's requirements. The task requirement typically includes semantic requirements based on tags and descriptions and statistical requirements based on RKME. Initially, the system executes a semantic filter across all learnwares, followed by a basic statistical search, which supports both singular and multiple learnware searches.
\begin{itemize}[leftmargin=*]
    \item \textbf{Single Learnware Search}: the system recommends learnware with similar data distribution. In details, given the learnware RKME specification $\s_l=\{\beta_{li},\z_{li}\}_{i=1}^{n_l}$ and user's RKME requirement $\s_u=\{\beta_{ui},\z_{ui}\}_{i=1}^{n_u}$, the system calculates the maximum mean discrepancy (MMD) distance between them: $\left\| \sum_{i=1}^{n_l} \beta_{li}k\left(\z_{li}, \cdot\right)-\sum_{j=1}^{n_u} \beta_{uj} k\left(\z_{uj}, \cdot\right)\right\|_{\mathcal{H}}^2$. The system transforms distances to scores and recommends learnwares with the score surpassing a preset threshold to the user.
    \item \textbf{Multiple Learnware Search}: When a single learnware inadequately addresses the user's task, the system attempts to assemble a combination of learnwares for user's task. This involves using a weighted combination of data distributions from the learnwares to approximate the user's task data distribution. The system employs two strategies based on single learnware search outcomes. The first strategy calculates the mixture weights for filtered learnwares and return learnwares with high weights. The second strategy, a more efficient approach, incrementally adds learnwares that significantly reduce the distribution distance.
\end{itemize}

{\bf System reuse algorithms.} Upon identifying and providing users with pertinent learnwares, the system provides various basic methods for learnware reuse. These methods allow users to effectively apply the learnwares to their tasks, thereby eliminating the need to develop models from scratch. There are two main categories:
\begin{itemize}[leftmargin=*]
    \item \textbf{Data-free reusers: reuse learnwares directly}:
        \begin{itemize}
            \item Average Ensemble~\citep{Zhou2012}. the \texttt{AveragingReuser} uniformly averages learnwares prediction.

            \item Job Selector~\citep{Wu:Xu:Liu:Zhou2023}: the \texttt{JobSelectorReuser} trains a multi-class classifier to identify the appropriate learnware for each user data. The classifier is trained either on the reduced set samples of learnware specification or the data herding from the specification.

        \end{itemize}
    \item \textbf{Data-dependent reusers: reuse learnwares with minor labeled data}:
    \begin{itemize}
        \item Ensemble Pruning~\citep{Wu:He:Qian:Zhou2022}: the \texttt{EnsemblePruningReuser} selects a subset from a provided learnware list using multi-objective evolutionary algorithm and uses the \texttt{AveragingReuser} for average ensemble. The evolutionary algorithm optimizes validation error, margin distribution, and ensemble size concurrently.

        \item Feature Augmentation: the \texttt{FeatureAugmentReuser} enhances user task features by incorporating predictions from learnwares, subsequently training a simple model. For classification tasks, logistic regression is employed, while ridge regression is utilized for regression tasks.

    \end{itemize}
\end{itemize}

\subsection{Handling Tabular Learnwares with Heterogeneous Feature Spaces.}
{\bf Tabular specification world.} 
The specification, detailing the model's specialty and utility, is a crucial component of the learnware. Each \textit{specification island} encompasses all models that share the same functional space $\mathcal{F}: \mathcal{X} \mapsto \mathcal{Y} \text { w.r.t. obj.}$, where $\mathcal{X}$ represents the input domain, $\mathcal{Y}$ the output domain, and $\text{obj}$ the objective. Collectively, these specification islands constitute the \textit{specification world}~\citep{Zhou:Tan2024}.

When a user submits her task requirements, the system initially identifies a specific specification island that aligns with the input domain $\mathcal{X}$ and output domain $\mathcal{Y}$ of the user's task, followed by a thorough search among corresponding learnwares. While this approach is straightforward, it encounters two major challenges. First, all models associated with a particular specification island might still be irrelevant to the user's current task due to discrepancies in the internal patterns of the training data. Second, a suitable specification island may not exist, which is really common for tabular tasks due to highly structured yet flexible data and complicated feature semantics.

To address these challenges for tabular tasks, the system must leverage learnwares from different specification islands and broaden the search scope by merging these islands into a unified tabular specification world. The system initially proposes the solution for creating this tabular specification world, encompassing all potential tabular specification islands. This process operates under mild conditions, requiring no additional auxiliary data across varying feature spaces, and is capable of merging any tabular specification islands.

{\bf Recommend heterogeneous tabular learnwares.}
To recommend potentially useful tabular learnwares from the entire collection, the system must integrate different tabular specification islands by establishing relationships between them. Notably, each specification island corresponds to a unique feature space $\mathcal{X}$, characterized by specific feature descriptions. The key strategy for merging these islands involves leveraging their relationships through feature semantics to create a unified specification world, based on semantic embeddings. In particular, the system employs a heterogeneous tabular network~\citep{Wang:Sun2022} to develop this unified semantic embedding space. This network serves as the system's heterogeneous engine, generating new specifications for each tabular learnware. For a given tabular learnware with specifications $\{\beta_i,\z_i\}_{i=1}^n$, the new specification retains the coefficient $\beta_i$ and transforms the samples $\z_i$ using the system engine $F(\cdot)$, resulting in the new specification $\{\beta_i,F(\z_i)\}_{i=1}^n$. Once all tabular specification islands are merged into this unified world, all specifications reside in the same space, enabling the tabular learnware recommendation to encompass all tabular learnwares, rather than just those with matching input and output domains.

{\bf Reuse heterogeneous tabular learnwares.} Upon receiving the heterogeneous learnware recommended by the system, users are still unable to apply it directly to their tasks due to discrepancies in input domain $\mathcal{X}$ and output domain $\mathcal{Y}$. Nevertheless, the system facilitates the reuse of heterogeneous learnware through a three-step process: 1) aligning the input domain, 2) predicting with the learnware, and 3) aligning the output domain.

In the first step, input alignment, only the RKME specification of the recommended learnware and the RKME requirement are utilized, ensuring the user's original data remains confidential. This step involves transforming the feature space based on a reduced set of learnware specifications, denoted as $\s_l=\{\beta_{li},\z_{li}\}_{i=1}^{n_l}$, and a corresponding set of user requirements, $\s_u=\{\beta_{ui},\z_{ui}\}_{i=1}^{n_u}$. The transformation, represented by $\phi$, minimizes the distance between two distributions: the learnware specification $\s_l=\{\beta_{li},\z_{li}\}_{i=1}^{n_l}$ and the projected user requirement $\s_u^{\text{proj}}=\{\beta_{ui},\phi(\z_{ui})\}_{i=1}^{n_u}$. Utilizing the maximum mean discrepancy (MMD) distance, $\phi$ is derived from the following optimization problem, which is solved using gradient descent:
$
    \min _{\phi}\left\|\sum_{i=1}^{n_u} \beta_{ui} k\left(\phi(\z_{ui}), \cdot\right)-\sum_{j=1}^{n_l} \beta_{lj} k\left(\z_{lj}, \cdot\right)\right\|_{\mathcal{H}}^{2}
$.
With $\phi$ generated, the user's data is transformed to align with the learnware's input domain, enabling prediction. However, since the output domain of the user's task might differ from that of the learnware, output alignment is required. This process involves using a small amount of labeled data and is conducted by augmenting features, i.e., incorporating learnware predictions as additional features to train a simple model, such as logistic regression for classification tasks and ridge regression for regression tasks.

\section{Experimental Evaluation}
\label{sec:evaluation}

In this chapter, we build various types of experimental scenarios and conduct extensive empirical study to evaluate the baseline algorithms for specification generation, learnware identification, and reuse on tabular, image, and text data.

\subsection{Experiments on Tabular Data}
On various tabular datasets, we initially evaluate the performance of identifying and reusing learnwares from the learnware market that share the same feature space as the user's tasks. Additionally, since tabular tasks often come from heterogeneous feature spaces, we also assess the identification and reuse of learnwares from different feature spaces. 

{\bf Settings.}
Our study utilize three public datasets in the field of sales forecasting: Predict Future Sales (PFS)~\citep{pfsdata}, M5 Forecasting (M5)~\citep{Makridakis:Spiliotis:Assimakopoulos2022}, and Corporacion~\citep{Corporaciondata}. To enrich the data, we apply diverse feature engineering methods to these datasets. Then we divide each dataset by store and further split the data for each store into training and test sets. A LightGBM is trained on each Corporacion and PFS training set, while the test sets and M5 datasets are reversed to construct user tasks. This results in an experimental market consisting of 265 learnwares, encompassing five types of feature spaces and two types of label spaces. All these learnwares have been uploaded to the learnware dock system.

\subsubsection{Homogeneous Cases}

In the homogeneous cases, the 53 stores within the PFS dataset function as 53 individual users. Each store utilizes its own test data as user data and applies the same feature engineering approach used in the learnware market. These users could subsequently search for homogeneous learnwares within the market that possessed the same feature spaces as their tasks.

\begin{figure}[!h]
    \begin{minipage}[c]{0.4\linewidth}
    \centering
    \renewcommand\arraystretch{1.3}
    \tabcolsep=0.1cm
    \begin{tabular}{c|c}
        \hline
        \textbf{Method} & \makebox[0.2\textwidth][c]{\textbf{Loss}}\\
        \hline
        Mean in Market & $0.897$\\
        \hline
        Best in Market & $0.756$\\
        \hline
        Top-1 & $0.830$\\
        \hline
        JobSelector & $0.848$\\
        \hline
        AverageEnsemble & $0.816$\\
        \hline
    \end{tabular}
    \end{minipage}
    \begin{minipage}[c]{0.55\linewidth}
    \centering
    \includegraphics[width=0.99\textwidth]{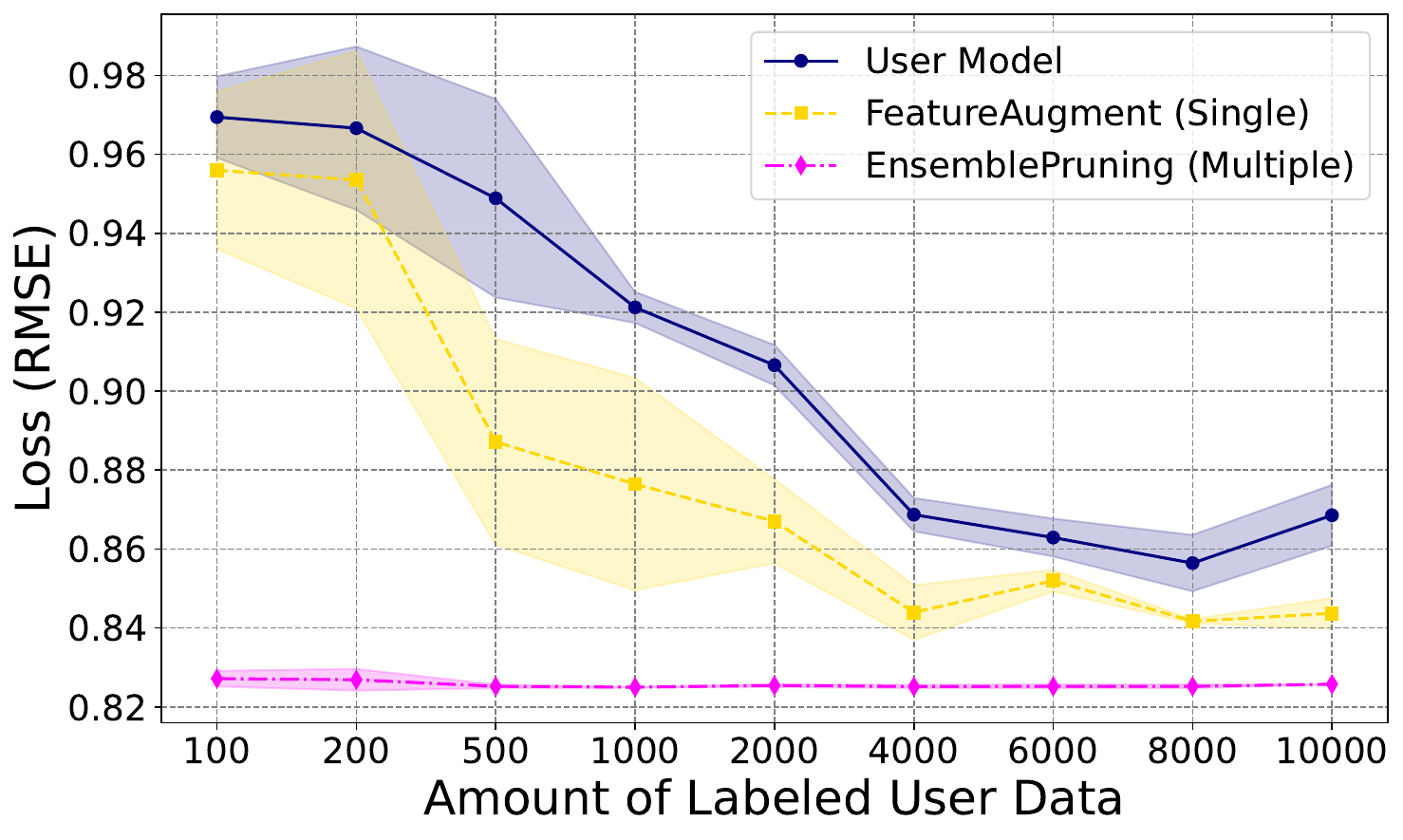}
    \end{minipage}
    \caption{Average performance of homogeneous table experiments. The table on the left displays the results obtained when the user has no labeled data available. The figure on the right showcases the results for different amounts of labeled data provided by the user; for each user, we conducted multiple experiments repeatedly and calculated the mean and standard deviation of the losses; the average losses over all users are illustrated in the figure.}
    \label{fig:table_homo_experiment}
\end{figure}

{\bf Results.}
We conduct a comparison among different baseline algorithms when the users have no labeled data or limited amounts of labeled data. Top-1 reuser is a baseline method that directly uses the best single learnware returned by the searcher, while other methods are different reusers introduced in Section~\ref{subsec:algorithm}. The average losses over all users are illustrated in Fig.~\ref{fig:table_homo_experiment}. The left table shows that data-free methods are much better than random choosing and deploying one learnware from the market.
The right figure illustrates that when users have limited training data, identifying and reusing single or multiple learnwares yields superior performance compared to user's self-trained models. 

\subsubsection{Heterogeneous Cases}

Based on the similarity of tasks between the market's learnwares and the users, the heterogeneous cases can be further categorized into different feature engineering and different task scenarios.

{\bf Different feature engineering scenarios.}
We consider the 41 stores within the PFS dataset as users, generating their user data using a unique feature engineering approach that differ from the methods employed by the learnwares in the market. As a result, while some learnwares in the market are also designed for the PFS dataset, the feature spaces do not align exactly. 

\begin{figure}[h]
    \begin{minipage}[c]{0.4\linewidth}
    \centering
    \renewcommand\arraystretch{1.3}
    \tabcolsep=0.1cm
    \begin{tabular}{c|c}
        \hline
        \textbf{Method} & \makebox[0.2\textwidth][c]{\textbf{Loss}} \\
        \hline
        Mean in Market & $1.149$\\
        \hline
        Best in Market & $1.038$\\
        \hline
        Top-1 & $1.075$\\
        \hline
        AverageEnsemble & $1.064$\\
        \hline
    \end{tabular}
    \end{minipage}
    \begin{minipage}[c]{0.55\linewidth}
    \centering
    \includegraphics[width=0.99\textwidth]{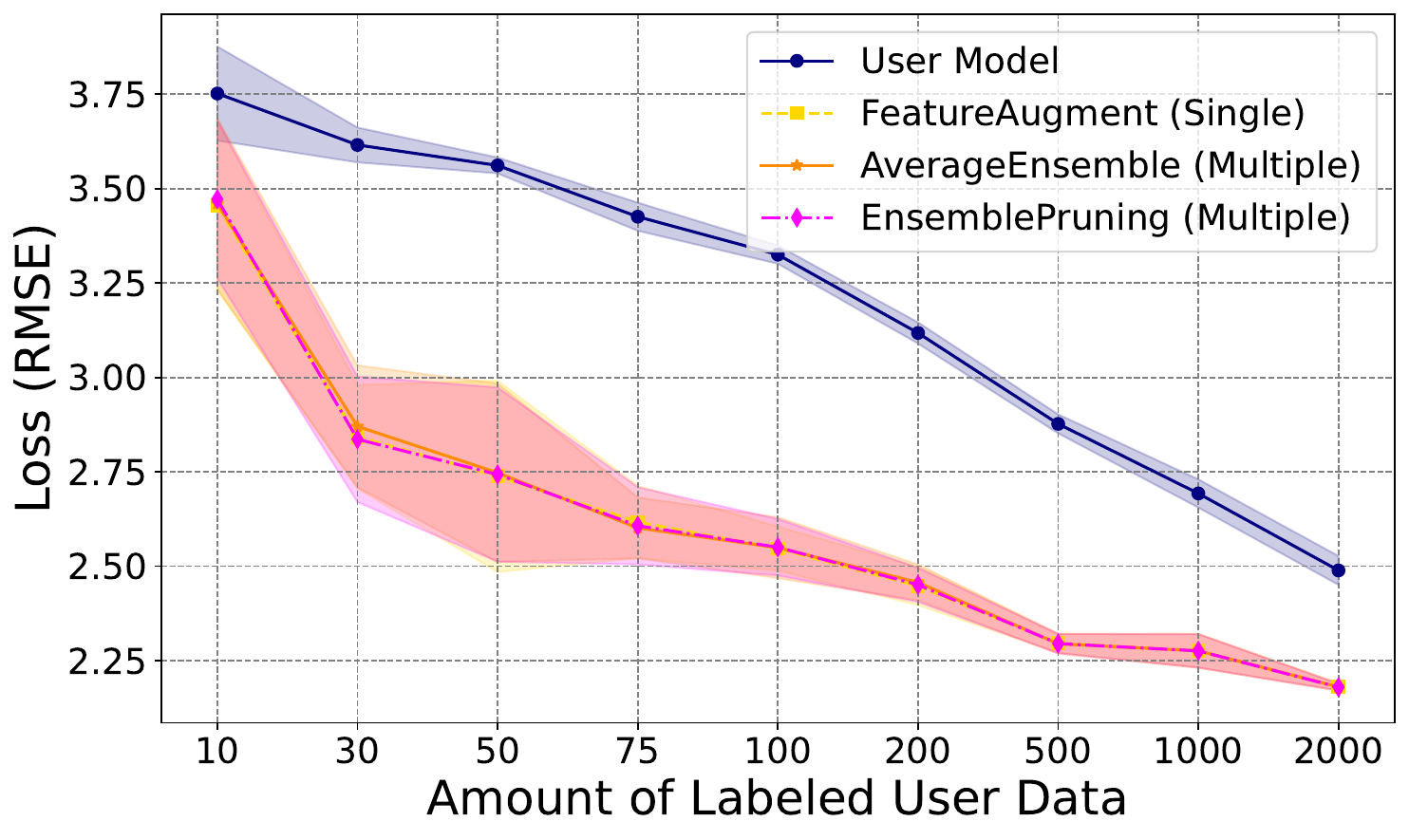}
    \end{minipage}
    \caption{Average performance of heterogeneous table experiments. The table in the left displays the results of different feature engineering experiments. The figure in the right shows the curves of different task experiments.}
    \label{fig:table_hetero_experiment}
\end{figure}

In this experimental setup, we examine various data-free reusers. The results displayed on the left side of Fig.~\ref{fig:table_hetero_experiment} indicate that even when users lack labeled data, the market exhibits strong performance, particularly with the AverageEnsemble method that reuses multiple learnwares.

{\bf Different task scenarios.}
We employ three distinct feature engineering methods on all the ten stores from the M5 dataset, resulting in a total of 30 users. Although the overall task of sales forecasting aligns with the tasks addressed by the learnwares in the market, there are no learnwares specifically designed to satisfy the M5 sales forecasting requirements. 

In the right side of Fig.~\ref{fig:table_hetero_experiment}, we present the loss curves for the user's self-trained model and several learnware reuse methods. It is evident that heterogeneous learnwares prove beneficial with a limited amount of the user's labeled data, facilitating better alignment with the user's specific task. 

\subsection{Experiments on Image and Text Data}
Second, we assess our system on image datasets. It is worth noting that images of different sizes could be standardized through resizing, eliminating the need to consider heterogeneous feature cases.

\begin{figure}[h]
    \begin{minipage}[c]{0.4\linewidth}
    \centering
    \renewcommand\arraystretch{1.3}
    \tabcolsep=0.1cm
    \begin{tabular}{c|c}
        \hline
        \textbf{Method} & \makebox[0.2\textwidth][c]{\textbf{Loss}}\\
        \hline
        Mean in Market & $0.655$\\
        \hline
        Best in Market & $0.304$\\
        \hline
        Top-1 & $0.406$\\
        \hline
        JobSelector & $0.406$\\
        \hline
        AverageEnsemble & $0.310$\\
        \hline
    \end{tabular}
    \end{minipage}
    \begin{minipage}[c]{0.55\linewidth}
    \centering
    \includegraphics[width=0.99\textwidth]{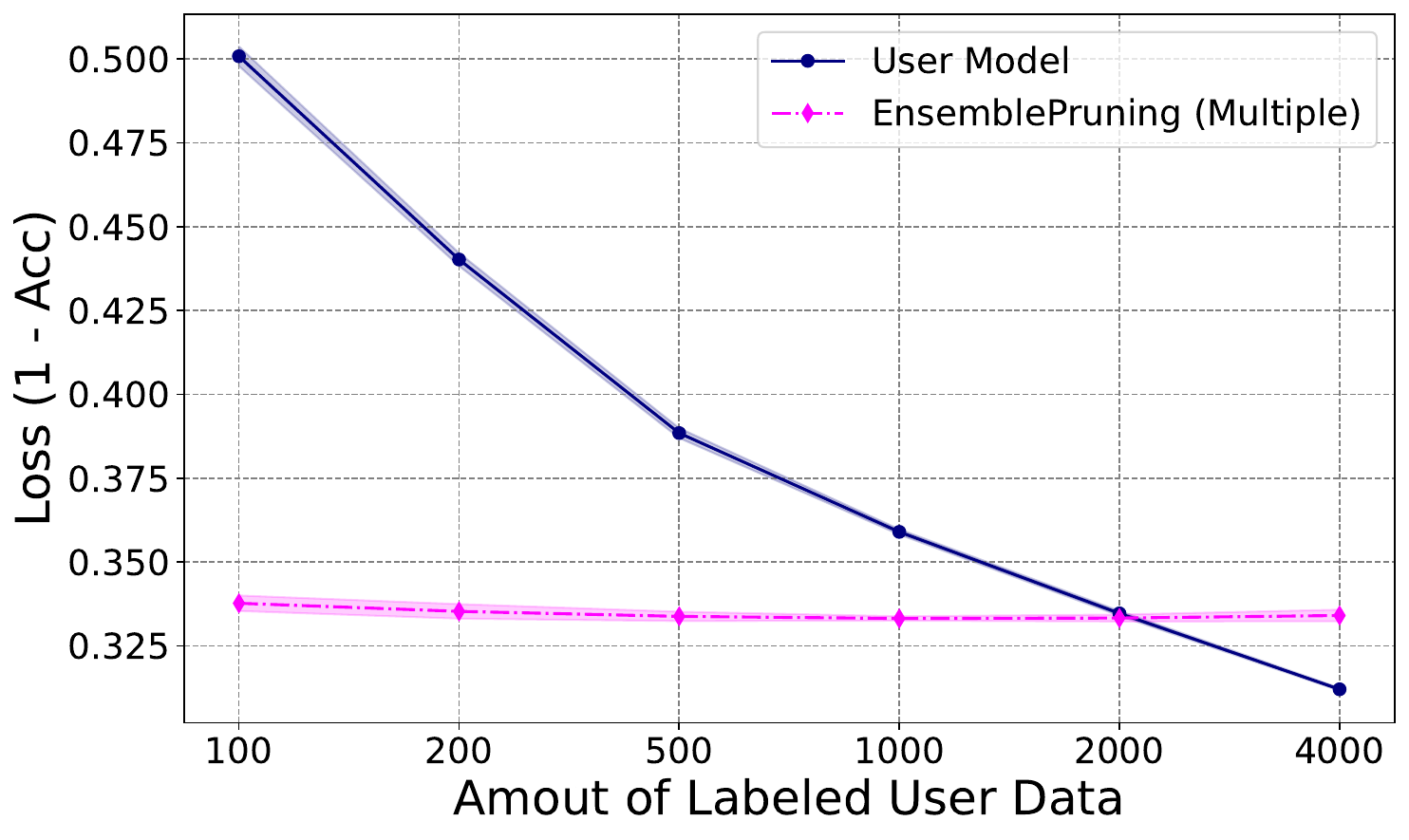}
    \end{minipage}
    \caption{Average performance of image experiments. The table on the left displays the results obtained when the user has no labeled data available. The figure on the right showcases the results for different amounts of labeled data provided by the user.}
    \label{fig:image_experiment}
\end{figure}

{\bf Settings.}
We choose the famous image classification dataset CIFAR-10~\citep{Krizhevsky2009}, which consists of 60000 32x32 color images in 10 classes. A total of 50 learnwares were uploaded: each learnware contains a convolutional neural network trained on an unbalanced subset that includes 12000 samples from four categories with a sampling ratio of $0.4:0.4:0.1:0.1$. 
A total of 100 user tasks are tested and each user task consists of 3000 samples of CIFAR-10 with six categories with a sampling ratio of $0.3:0.3:0.1:0.1:0.1:0.1$.

{\bf Results.} We assess the average performance of various methods using 1 - Accuracy as the loss metric. Fig.~\ref{fig:image_experiment} shows that when users face a scarcity of labeled data or possess only a limited amount of it (less than 2000 instances), leveraging the learnware market can yield good performances.

Finally, we evaluate our system on text datasets. Text data naturally exhibit feature heterogeneity, but this issue can be addressed by applying a sentence embedding extractor.

{\bf Settings.}
We conduct experiments on the well-known text classification dataset: 20-newsgroup~\citep{Joachims1997}, which consists approximately 20000 newsgroup documents partitioned across 20 different newsgroups.
Similar to the image experiments, a total of 50 learnwares were uploaded. Each learnware is trained on a subset that includes only half of the samples from three superclasses and the model in it is a tf-idf feature extractor combined with a naive Bayes classifier. We define 10 user tasks, and each of them encompasses two superclasses.

\begin{figure}[h]
    \begin{minipage}[c]{0.4\linewidth}
    \centering
    \renewcommand\arraystretch{1.3}
    \tabcolsep=0.1cm
    \begin{tabular}{c|c}
        \hline
        \textbf{Method} & \makebox[0.2\textwidth][c]{\textbf{Loss}}\\
        \hline
        Mean in Market & $0.493$\\
        \hline
        Best in Market & $0.141$\\
        \hline
        Top-1 & $0.154$\\
        \hline
        JobSelector & $0.155$\\
        \hline
        AverageEnsemble & $0.138$\\
        \hline
    \end{tabular}
    \end{minipage}
    \begin{minipage}[c]{0.55\linewidth}
    \centering
    \includegraphics[width=0.99\textwidth]{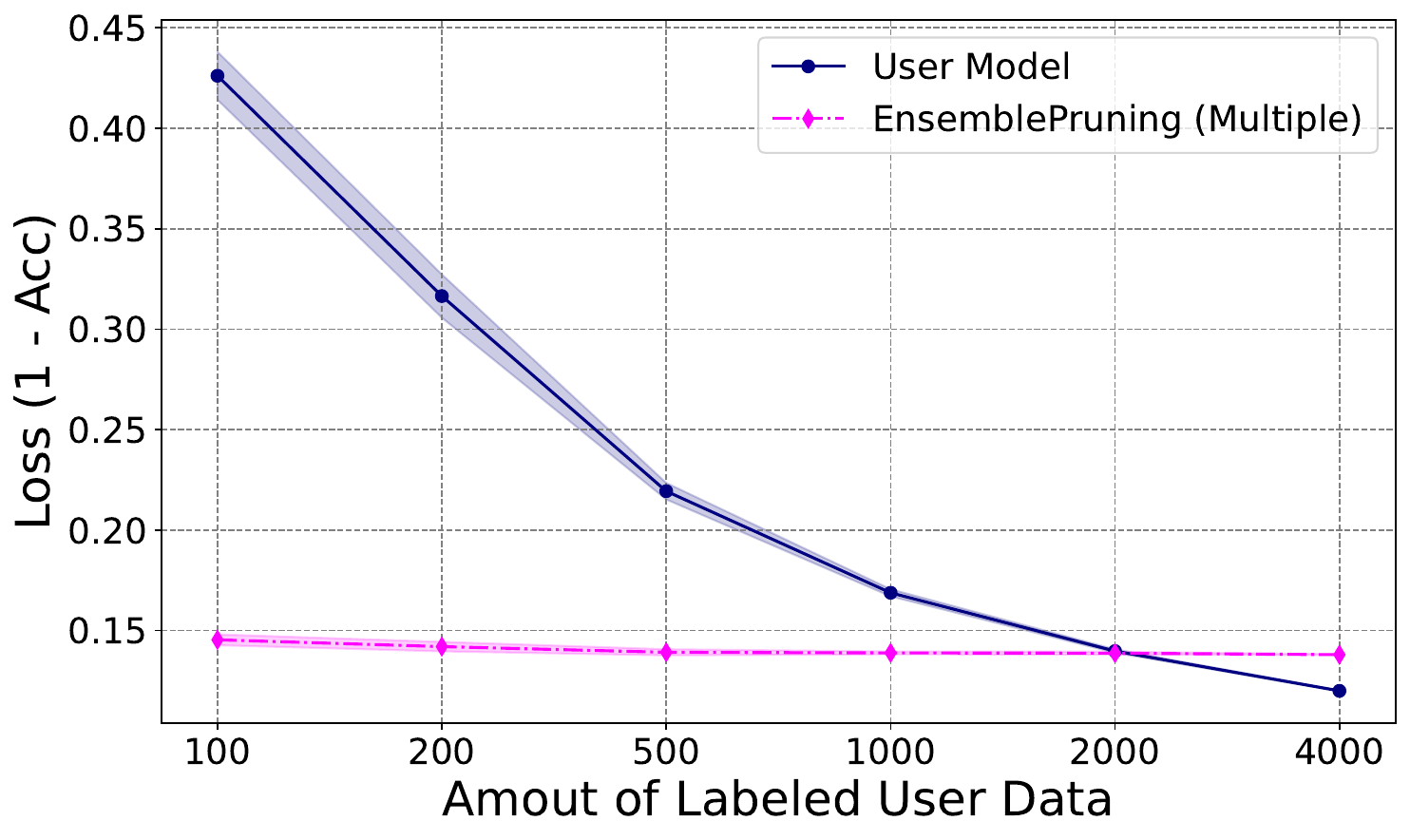}
    \end{minipage}
    \caption{Average performance of text experiments. The table on the left displays the results obtained when the user has no labeled data available. The figure on the right showcases the results for different amounts of labeled data provided by the user.}
    \label{fig:text_experiment}
\end{figure}

{\bf Results.}
The results are depicted in Figure~\ref{fig:text_experiment}. Similarly, even when no labeled data is provided, the performance achieved through learnware identification and reuse can match that of the best learnware in the market. Additionally, utilizing the learnware market allows for a reduction of approximately 2000 samples compared to training models from scratch.

\section{Related Work}
\label{sec:related}

In the following, we first show the difference of primary focuses between learnware paradigm and classic fields that utilize the given source tasks or models to solve specific target tasks.

{\bf Utilizing given source task(s) or model(s).} There are several fields focusing on solving specific target tasks assuming one or several helpful source task(s) or model(s) have been given. 
Domain adaptation~\citep{Ben-David:Blitzer:Crammer:Pereira2006} and transfer learning~\citep{Pan:Yang2009} aim to transfer knowledge from a source domain to a target domain, and usually assumes that the raw source data is available~\citep{Pan:Tsang:Kwok:Yang2010,Zhuang:Qi:Duan:Xi2021}. To relax the requirements for source data, several topics focus on adapting the models from source domain to the target domain, like source-free domain adaptation~\citep{Yang:Wang:Wang:Jui2022}, hypothesis transfer learning~\citep{Kuzborskij:Orabona2013}, model reuse~\citep{Zhao:Cai:Zhou2020}, domain adaptation with auxiliary classifiers~\citep{Duan:Tsang:Xu:Chua2009,Li:Tsang:Zhou2013}, etc. 
The learnware paradigm highly differs from these fields because the learnware paradigm aims to build a learnware dock system to leverage numerous high-quality models from the community in a unified and privacy-preserving way to solve unplanned new tasks, in which a big challenge is to efficiently identify the most helpful learnware(s) for the unplanned new task to reuse.

Subsequent to the proposal of learnware paradigm, recently several related research streams have emerged. In the following, we will introduce the recent related works, mostly focusing on specific aspects without a unified comprehensive architecture design. 

{\bf Model platforms.} Currently, with the expanding application of machine learning models across various scenarios, there has been a significant growth in the development of model platforms/pools/hubs, with Hugging Face platform~\footnote{https://huggingface.co/models} being the most popular containing nearly half a million models. As the learnware paradigm envisions, with the ever-increasing number of models, how to identify the most suitable models for a user task becomes more and more important and challenging.
These platforms generally operates as a git-based remote hosting service for managing submitted models like text and code, however, models correspond to functions realizing mappings from the input domain to the output domain. 
By leveraging various models in a unified way based on specification, the learnware paradigm fundamentally differs from model pools/hubs which provides passive hosting. As described above, by generating the \emph{specification} for each submitted model which can statistically and semantically capture and represent the specialty of various models, like RKME specification, identifying truly suitable model(s) effectively and efficiently becomes achievable. Without specifications, when facing a new task, it would be forced to examine all models and their combinations in the whole platform on user data to identify the most helpful ones, which is unacceptable in terms of data privacy and computational efficiency. In \bmwu{}, we propose a novel specification-based architecture which can leverage various models in a unified way. Instead of passive hosting, \bmwu{} can help users to easily obtain a deployable well-performing model suitable for their tasks with just a few lines of code, without extensive data and expert knowledge, while ensuring data privacy. 

Besides, there is a series of research to generally reduce the technical burden for model developers to deploy a shared model locally, like Infaas~\citep{Romero:Li:Yadwadkar:Kozyrakis2021}, Acumos~\citep{Zhao:Talasila:Jacobson2018}, etc, which mainly focus on the deployment phase and doesn't involve a huge platform system and the process of sufficient model identification. Recently, there are some works proposed to adopt the popular large language models (LLMs)~\citep{Brown:Mann:Ryder:Subbiah2020} to identify helpful models based on matching the natural language descriptions of each model in the platform and the user’s requirements, like HuggingGPT~\citep{Shen:Song:Tan:Li2023}. These works utilizes semantic information only, whereas in most cases accurate model identification can not be realized without statistical specification, and accurate statistical specification can enable models to be used beyond their original purposes.
There are some studies about assessing the reusability or transferability of pre-trained models~\citep{Nguyen:Hassner:Seeger:Archambeau2020,Ding:Wu:Zhou:Zhou2022,You:Liu:Zhang:Wang2022,Zhang:Huang:Ding:Zhan2023}. They assume raw labeled data of the user is available in the assessment procedure and generally utilize all pre-trained models to make the forward pass for all user data without considering data privacy. Besides, it is unaffordable to access their combinations on user data due to combinatorial explosion.

\section{Conclusion}

In this paper, we present \bmwu{}: a open-source learnware dock system providing foundational support for future research of learnware paradigm. \bmwu{} has been released with online service~\footnote{https://bmwu.cloud/}, and open-source repository~\footnote{https://www.gitlink.org.cn/beimingwu/beimingwu}. The \texttt{learnware} package, serving as the core engine of the system, has also been released online~\footnote{https://www.gitlink.org.cn/beimingwu/learnware}. 
In \bmwu, for the first time, we specify a unified learnware structure, and design an integrated architecture for the system and the core engine, supporting the entire process including the submitting, usability testing, organization, management, identification, deployment and reuse of learnwares. 
The system sets a solid foundation for future explorations in learnware-related algorithms, and lays the groundwork for accommodating and leveraging a vast array of learnwares and establishing a learnware ecosystem. 
Due to the vast complexity of real-world tasks and limited existing learnwares, currently \bmwu{} has limitations for numerous specific and unforeseen scenarios. However, the foundational architecture and implementations enable the system to continuously expand its knowledge base and improve its capabilities through the constant submission of learnwares and advancements in algorithmic research, and this continuous evolution of learnware dock system equips it with lifelong learning capability to tackle more varied user tasks.

\section*{Acknowledgement}

This work was supported by the National Science Foundation of China (62250069). We appreciate the support of Polixir team in various aspects like system deployment and industrial applications. We appreciate the contributions of Hao-Yu Shi and Xin-Yu Zhang for the support in image and text scenarios. We appreciate the contributions of Lan-Zhe Guo, Zi-Xuan Chen, Zhi Zhou and Yi-Xuan Jin for the help in early-stage prototype. We thank LAMDA faculties for their generous help and insightful discussions. We are also grateful for LAMDA members for their contributions in the system's internal testing phase.

\bibliographystyle{plainnat}
\bibliography{bmwu}

\end{document}